\documentclass[twocolumn,showpacs,aps]{revtex4}
\usepackage{graphics}

\begin{document}

\title{Characterization of the probabilistic traveling salesman problem}

\author{Neill E. Bowler}
\email{Neill.Bowler@physics.org}
\homepage{http://uk.geocities.com/neill_bowler}
\affiliation{Department of Physics, University of Warwick, Coventry,
CV4 7AL, England}
\affiliation{Met Office,  Maclean Building, Crowmarsh-Gifford, Oxfordshire OX10 8BB, England}

\author{Thomas M. A. Fink}
\email{tmf20@cam.ac.uk}
\homepage{http://www.tcm.phy.cam.ac.uk/~tmf20/}
\affiliation{CNRS UMR 144, Institut Curie, 75005 Paris, France}
\affiliation{Theory of Condensed Matter, Cavendish Laboratory,
Cambridge, CB3 0HE, England}

\author{Robin C. Ball}
\email{r.c.ball@warwick.ac.uk}
\homepage{http://www.phys.warwick.ac.uk/theory/}
\affiliation{Department of Physics, University of Warwick, Coventry,
CV4 7AL, England}

\date{\today}

\begin{abstract}
We show that Stochastic Annealing can be successfully applied to gain new
results on the 
Probabilistic Traveling Salesman Problem (PTSP).  The probabilistic ``traveling
salesman'' must decide on an a priori order in which to visit $n$ cities
(randomly distributed over a
unit square) \emph{before} learning that some cities can be omitted.  We find
the optimized average length of the pruned tour follows
$E(\bar{L}_{\mbox{pruned}})=\sqrt{np}(0.872-0.105p)f(np)$ where $p$ is the probability
of a city needing to be visited, and $f(np)\rightarrow 1$ as $np\rightarrow \infty $.
The average length of the a priori tour (before omitting any cities) is found
to follow $E\left( L_{\mbox{a priori}}\right) =\sqrt{\frac{n}{p}}\beta (p)$ 
where $\beta (p)=\frac{1}{1.25-0.82\, ln(p)}$ is measured for
$0.05\leq p\leq 0.6$. Scaling arguments and indirect measurements suggest that
$\beta (p)$ tends towards a constant for $p<0.03$.  Our
stochastic annealing algorithm is based on limited sampling of the pruned
tour lengths, exploiting the sampling error to provide the analogue of thermal
fluctuations in simulated (thermal) annealing.  The method has general
application to the optimization of functions whose cost to evaluate rises
with the precision required.
\end{abstract}

\maketitle

\section{Introduction}

Many real systems present problems of stochastic optimization. These include
communications networks, protein design \cite{fink1} and oil field models 
\cite{jonsbraten}, \ in all of which uncertainty plays a central role. We
will consider the case where the outcome $g(x,\omega )$ depends not only on
parameters $x$ to be chosen, \ but also on unknowns $\omega $. We can only
average with respect to these unknowns, aiming to find the `solution' $x$
which optimises the average outcome. Thus we seek to find $x\in X$ which
minimizes 
\begin{equation}
\bar{g}(x)=\int g(x,\omega )f(\omega )\,d\omega
\end{equation}
where $X$ is the solution space of the problem and $f(\omega )$ is the
probability distribution of the uncertain variables. 

Stochastic optimization was borne out of an idea by Robbins \& Monro \cite
{robbins}. They considered solving the problem of finding 
\begin{equation}
G(x)=\alpha
\end{equation}
where $G$ is some monotonic function of $x$ and $\alpha $ is a parameter. $G$
is not known directly, but can only be estimated. Their technique to solve
this problem is called stochastic approximation, \ and a number of variants
of this scheme have since been developed \cite
{benveniste,kushner1,l'ecuyer,kushner2}.

Stochastic optimization is a more general situation since the function to be
minimized may have many local minima. We may classify the techniques to
solve stochastic optimization problems into two classes - exact methods and
heuristics. Heuristics are more appropriate to NP-complete problems, and
these are the problems on which we focus in this paper. \ A number of
heuristics already exist to tackle stochastic optimization problems \cite
{gong,yan,devroye,yakowitz,andradottir}. Many of these are developments from
simulated annealing \cite{haddock,bulgak,alrefaei,alkhamis}, \ which has
been shown \cite{gutjahr} to solve stochastic optimization problems with
probability 1, provided $\bar{g}(x)$ can be estimated with precision greater
than $O(t^{-\gamma })$ for time step $t$, where $\gamma >1$. \ A number of
authors \cite{bulgak,alrefaei,alkhamis,gelfand} have used a modified
simulated annealing algorithm in which the acceptance probability is
modified to take some account of the precision of the estimates of $\bar{g}%
(x)$, \ and in these cases there are a number of convergence results \cite
{gelfand,alkhamis}.

Stochastic annealing \cite{fink1} is a modified simulated annealing
algorithm which differs from the above approaches in two key ways. Firstly
the noise present in estimates is positively exploited as mimicking thermal
noise in a slow cooling, \ as opposed to being regarded as something whose
influence should be minimised from the outset. Secondly, stochastic
annealing can be modified to give exact simulation of a thermal system.
Although this is not specifically ruled out by the earlier approaches, no
attempt has been made with them to satisfy this condition.

In stochastic annealing we estimate $\bar{g}(x)$ by taking $r$ repeated,
statistically independent, measurements of $g(x,\omega )$ each of which we
call an instance. For the implementation of stochastic annealing used in
this paper we accept all moves for which one estimate (based on $r$
instances) for a new state is more favourable than an equivalent estimate
for the old. This simple procedure does not exactly simulate a thermal
system, \ where the acceptance probabilities should obey 
\begin{equation}
\frac{P_{A\rightarrow B}}{P_{B\rightarrow A}}=e^{-\beta \Delta \mu }
\label{detailed balance}
\end{equation}
where $\beta =\frac{1}{k_{B}T}$ and $\Delta \mu $ is the exact difference in 
$\bar{g}(x)$ between states $A$ and $B$. However if we assume that our
estimate change is Gaussian distributed around $\bar{g}(x)$ with standard
deviation $\frac{\sigma }{\sqrt{r}}$ 
, where $r$ is the number of instances used for each estimate, then it
follows that the acceptance probability is \cite{fink1} 
\begin{equation}
P_{A\rightarrow B}^{G}=\frac{1}{2}\left[ 1-\text{erf}\left( \frac{\sqrt{r}%
\Delta \mu }{\sqrt{2}\sigma }\right) \right] .  \label{basicaccept}
\end{equation}
The approximation to a thermal acceptance rule is then quite good since 
\begin{equation}
\begin{array}{lll}
ln\left( \frac{P_{A\rightarrow B}}{P_{B\rightarrow A}}\right) & =ln\left( 
\frac{1-{\text{erf}}\left( \frac{\sqrt{r}\Delta \mu }{\sqrt{2}\sigma }%
\right) }{1+{\text{erf}}\left( \frac{\sqrt{r}\Delta \mu }{\sqrt{2}\sigma }%
\right) }\right) &  \\ 
& \simeq -\beta _{G}\Delta \mu -\frac{4-\pi }{48}(\beta _{G}\Delta \mu
)^{3}-\ldots \label{detailed 2} & 
\end{array}
\end{equation}
where 
\begin{equation}
\beta _{G}=\frac{\sqrt{8r}}{\sqrt{\pi }\sigma }  \label{temperature}
\end{equation}
identifies the equivalent effective temperature. The small coefficient ($%
\simeq 0.02$) of the cubic term in eq. \ref{detailed 2} makes this a rather
good approximation to true thermal selection.

Increasing sample size $r$ means that we are more stringent about not
accepting moves that are unfavourable, equivalent to lowering the
temperature, which is quantified by eq.\ \ref{temperature} for the Gaussian
case. As with standard simulated annealing \cite{rees,nulton,salamon1}, the
question of precisely what cooling schedule to use remains something of an
art.

\section{Probabilistic Traveling Salesman Problem (PTSP)}

We adopt the PTSP as a good test-bed amongst stochastic optimization
problems, in much the same way as the TSP has been considered a standard
amongst deterministic optimization problems. The PTSP falls into the class
of NP-complete problems \cite{bertsimas4}, and the TSP is a subset of the
PTSP.

The original traveling salesman problem (TSP) is to find the shortest tour
around $n$ cities, in which each city is visited once. For small numbers of
cities this is an easy task, but the problem is NP-complete: it is believed
for large $n$ that there is no algorithm which can solve the problem in a
time polynomial in $n$. \ Consideration of the traveling salesman problem
began with Beardwood \emph{et al.\ }\cite{beardwood}. They showed that in
the limit of large numbers of cities which are randomly distributed on the
unit square, the optimal tour length ($L_{\text{TSP}}$) follows \cite
{steele1} 
\begin{equation}
E(L_{\text{TSP}})=\beta _{\text{TSP}}\sqrt{n}+\alpha
\end{equation}
where $\beta _{\text{TSP}}$ and $\alpha $ are constants. Here and below $%
E(L) $ denotes the quantity $L$ averaged, after optimization, with respect to
different city positions, randomly placed on the unit square. Numerical
simulation \cite{lee} gives $\beta _{\text{TSP}}=0.7211(3)$ and $\alpha
=0.604(5)$ as estimates when $n\geq 50$. Significant divergence from this
behaviour is found for $n\leq 10$, but numerical estimates can be found
quickly (see appendix).

The probabilistic traveling salesman problem (PTSP), introduced by Jaillet 
\cite{jaillet1,jaillet2}, is an extension of the traveling salesman problem
to optimization in the face of unknown data. Whereas all of the cities in
the TSP must be visited once, in the PTSP each city only needs to be visited
with some probability, $p$. One first decides upon the order in which the
cities are to be visited, the `a priori' tour. Subsequently, it is revealed
which cities need to be visited, and those which do not need to be visited
are skipped to leave a `pruned tour'. The order in which the cities are to
be visited is preserved when pruning superfluous cities. The objective is to
chose an a priori tour which minimizes the average length of the pruned
tour. It is clear from figure \ref{spikytour} that near optimal a priori
tours may appear very different for different values of $p$.

In our terminology, the average pruned tour length is averaged over all
possible instances of which cities require to be visited. This was given by
Jaillet as \cite{jaillet1} 
\begin{equation}
\bar{L}_{\text{pruned}}=\sum_{q=0}^{n-2}\ p^{2}(1-p)^{q}\ L_{t}^{(q)}
\label{prunedlength}
\end{equation}
where 
\begin{equation}
L_{t}^{(q)}=\sum_{j=1}^{n}d(j,1+(j+q)_{\mbox{mod} \, n})
\end{equation}
is the sum of the distances between each city and its $(q+1)^{th}$ following
city on the a priori tour, and the factors $p^{2}(1-p)^{q}$ in the preceding
equation simply give the probability that any particular span skipping $q$
cities occurs in the pruned tour. \ Jaillet's closed form expression\ for
the average pruned tour length renders the PTSP to some extent accessible as
a standard (but still NP complete) optimization problem, and provides some
check on the PTSP results by stochastic optimization methods.

\begin{figure}[tbp]
\resizebox{8.6cm}{5.5cm}{\includegraphics{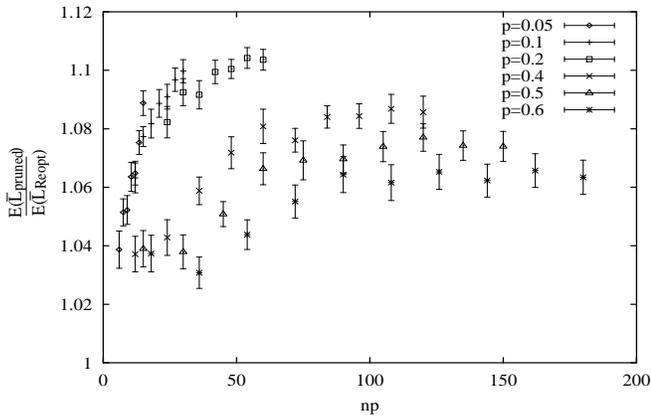}}
\caption{The expected pruned tour length divided by the expected
re-optimized tour length. This indicates the improvement one would expect
from re-optimization.}
\label{prunedgraph2}
\end{figure}

It has been conjectured \cite{bertsimas4} that, in the limit of large $n$,
the PTSP strategy is as good as constructing a TSP tour on the cities
requiring a visit, the re-optimization strategy. This would mean that
\begin{equation}
\lim_{n\rightarrow \infty }\left( \frac{E\left( \bar{L}_{\text{pruned}%
}\right) }{\sqrt{np}}\right) =\beta _{\text{TSP}},
\end{equation}
where $E(\bar{L}_{\text{pruned}})$ is the pruned tour length further
averaged over city positions after optimisation, which we will refer to as
the expected pruned tour length. \ Figure \ \ref{prunedgraph2} shows the
expected pruned tour length divided by the expected re-optimized tour
length. Since this quantity is tending towards a value significantly greater
than 1 for $p<1$ it demonstrates that the PTSP strategy can be worse than
the re-optimization strategy. \ Jaillet \cite{jaillet1} and Bertsimas \emph{%
et al.\ }\cite{bertsimas2} have also shown that there is a limit to how much
worse it can be, \ with 
\begin{equation}
\lim_{n\rightarrow \infty }\left( \frac{E\left( \bar{L}_{\text{pruned}%
}\right) }{\sqrt{np}}\right) =\beta _{\text{pruned}}(p)
\label{pruned variation}
\end{equation}
where 
\begin{equation}
\beta _{\text{TSP}}\leq \beta _{\text{pruned}}(p)\leq Min(0.9212,\frac{\beta
_{\text{TSP}}}{\sqrt{p}}).  \label{Beta limits}
\end{equation}

One attempt to solve the PTSP using an exact method was taken by Laporte 
\emph{et al.\ }\cite{laporte} who introduced the use of integer linear
stochastic programming. Although use of algorithms which may exactly solve
the PTSP are useful, they are always very limited in the size of problem
which may be attempted. Furthermore, the stochastic programming algorithm
even fails to solve the PTSP on certain occasions, thus the accuracy of any
statistics that would be generated using this method is dubious.

Three studies have used heuristics to solve the PTSP \cite
{bertsimas1,rossi,bertsimas5}. None of these studies used global search
heuristics, and all were very restricted in the problem size attempted due
to computational cost. The evaluation of a move for the PTSP using equation 
\ref{prunedlength} involves the computation of  $O(n^{2})$ terms compared to 
$O(1)$ computations to evaluate a move in the TSP. Thus, to solve a 100 city
problem for the PTSP would take $O\left( 10,000\right) $ times longer than
it would to solve a 100 city problem for the TSP. It should be noted,
however, that it is only possible to make this comparison due to the
relative simplicity of the PTSP. For many more stochastic optimization
problems, standard optimization techniques are simply not applicable.

\section{Form Of The Optimal Tour \& Scaling Arguments}

\begin{figure*}[t]
\resizebox{17.2cm}{8.5cm}{\includegraphics{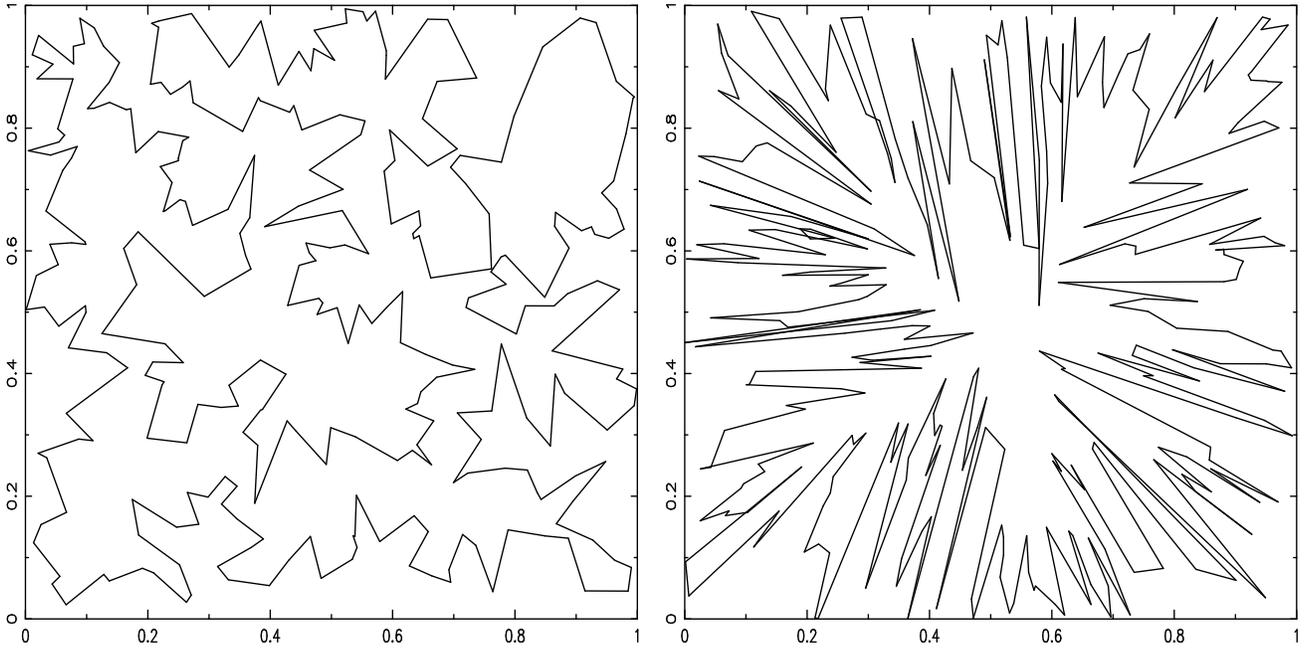}}
\caption{Typical near optimal a priori PTSP tours with $n=300$ for $p=0.5$%
(left) and $p=0.1$(right), respectively.}
\label{spikytour}
\end{figure*}

Optimal a priori PTSP tours for small $p$, as exemplified in figure \ref
{spikytour} for $p=0.1$, resemble an ``angular sort'' - where cities are
ordered by their angle with respect to the centre of the square. Bertsimas 
\cite{bertsimas1} proposed that an angular sort be optimal as $p\rightarrow
0 $, but we can show this to be false by comparison to a space-filling curve
algorithm which is generally superior as $n\rightarrow \infty $. Such an
algorithm was introduced by Bartholdi \emph{et al.\ }\cite{bartholdi} using
a technique based on a Sierpinski curve.

For the angular sort with $np\gg 1,$ the probability of two cities being
nearest neighbours on the pruned tour will be vanishingly small for cities
which are separated from each other by a large angle on the a priori tour.
This means that only cities that are separated by a small angle contribute
significantly to eq. \ref{prunedlength}. Thus for an $n$ city tour chosen by
angular sort, we may approximate 
\begin{equation}
L_{t}^{(q)}\simeq L_{o}n  \label{Angulardef}
\end{equation}
where $L_{o}$ is some fraction of the side of the unit square, since cities
which are sorted with respect to angle will be unsorted with respect to
radial distance. This leads to 
\begin{equation}
E(\bar{L}_{\text{ang}})\simeq L_{o}np^{2}\sum_{q=0}^{n-2}(1-p)^{q}.
\end{equation}
For $np\gg 1$ and $p\ll 1$, we then find that the angular sort yields 
\begin{equation}
E(\bar{L}_{\text{ang}})\rightarrow L_{o}np.
\end{equation}
By contrast it has been shown\cite{bertsimas2} that 
\begin{equation}
\frac{E(\bar{L}_{\tau _{\text{sf}}})}{E(\bar{L}_{\text{Reopt}})}=C
\end{equation}
with probability 1, where $E(\bar{L}_{\tau _{\text{sf}}})$ is the expected
length of a tour generated by a heuristic based on the Sierpinski curve and $%
E(\bar{L}_{\text{Reopt}})$ is the expected length for the re-optimization
strategy. Using previous computational results \cite{bertsimas2,lee}, we
estimate $C\simeq 1.33$, which is worse than we achieve using stochastic
annealing. Hence, $E(\bar{L}_{\tau _{\text{sf}}})$ is given by 
\begin{equation}
E(\bar{L}_{\tau _{\text{sf}}})=O(\sqrt{np})
\end{equation}
which leads to 
\begin{equation}
\frac{E(\bar{L}_{\tau _{\text{sf}}})}{E(\bar{L}_{\text{ang}})}=O(\frac{1}{%
\sqrt{np}}).
\end{equation}
So for large enough $np$, the angular sort is not optimal.

From inspection of near-optimal PTSP tours such as fig.\ \ref{spikytour}, we
propose that the tour behaves differently on different length scales; the
tour being TSP-like at larger length scales, but resembling a locally
directed sort at smaller length scales. We may construct such a tour and use
scaling arguments to analyse both the pruned and a priori lengths of the
optimal tour. \ Consider dividing the unit square into a series of `blobs',
each blob containing $1/p$ cities so that of order one city requires a
visit. The number of such blobs is given by 
\begin{equation}
N\simeq np
\end{equation}
and for these to approximately cover the unit square their typical linear
dimension $\xi $ must obey 
\begin{equation}
N\xi ^{2}\sim 1.
\end{equation}
Since each blob is visited of order once by a pruned tour, we can estimate
the expected pruned tour length to be 
\begin{equation}
E(\bar{L}_{\text{pruned}})\sim N\xi \sim \sqrt{np}  \label{prunedscaling}
\end{equation}
which we will see below is verified numerically. We can similarly estimate
the a priori tour length to be $n$ times the distance between two cities in
the same blob. Thus, the expected a priori tour length is 
\begin{equation}
E\left( L_{\text{a priori}}\right) \sim n\xi \sim \sqrt{\frac{n}{p}}
\label{a priori scaling}
\end{equation}
which is more difficult to confirm numerically.

\section{Computational Results For The PTSP}

We have investigated near optimal PTSP tours for a range of different
numbers of cities, and various values of $p$. We used stochastic annealing
with effective temperatures in the range $kT=0.07-0.01$, corresponding to
sample sizes in the range $r=2-500$. Between $10$ and $80$ different random
city configurations were optimized ($80$ configurations of 30 cities, $40$
configurations of 60 cities, $20$ configurations of 90 cities and $10$
configurations for $n\geq 120$ cities).

\begin{figure}[tbp]
\resizebox{8.6cm}{5.5cm}{\includegraphics{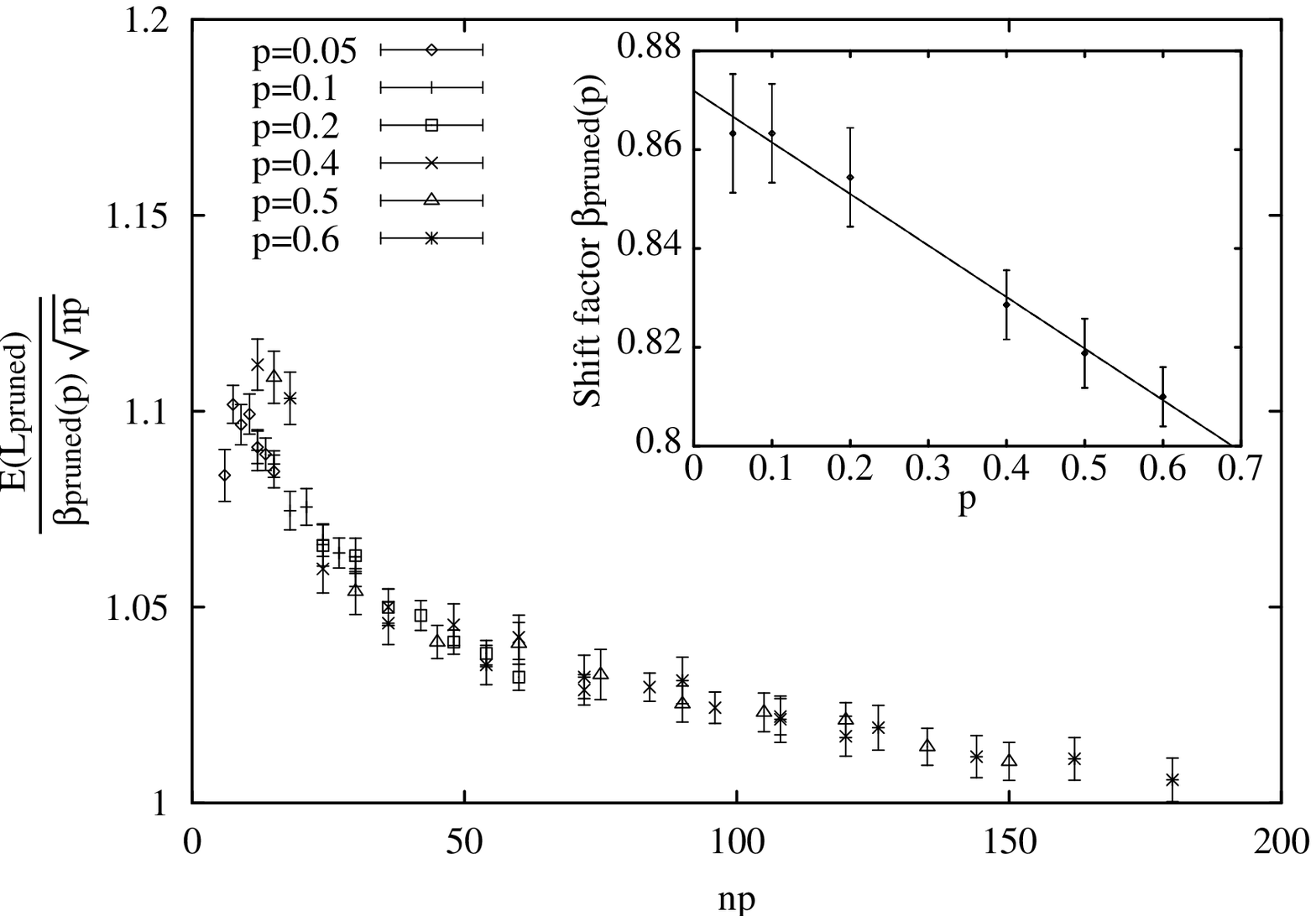}}
\caption{The master curve for the pruned tour length divided by $\protect%
\beta _{\mbox{pruned}}(p)\protect\sqrt{np}$. The data follows a smooth curve
for $n>30$, and the shift factors follow a linear relationship, suggesting
that $\frac{E\left( \bar{L}_{\mbox{pruned}}\right) }{\protect\sqrt{np}%
(0.872-0.105p)}=f(np)$. \ Three points with $n=30$ can be seen to fit less
well (here and also in fig. 4), \ showing breakdown of the master curve at
small $n$.}
\label{prunedgraph}
\end{figure}

Figure \ref{prunedgraph} shows a master curve for the expected pruned tour
length divided by $\sqrt{np}$. The shift factors have a linear fit and the
data are consistent with 
\begin{equation}  \label{prunedlaw}
\frac{E(\bar{L}_{\text{pruned}})}{\sqrt{np}(a-bp)}=f(np)
\end{equation}
for $n\gg 1$, where $a=0.872\pm 0.002$, $b=0.105\pm 0.005$ and $%
f(np)\rightarrow 1$ for large $np$. The shift factors indicate that the PTSP
strategy can be no more than $\frac{0.872}{0.767}-1=14 \left( \pm 1 \right)$ worse
than the re-optimization strategy.

The master curve for the a priori tour length is shown in fig.\ \ref{a
priori graph}. Our scaling arguments predict that the shift factors $\beta _{%
\text{a priori}}(p)$ should tend towards a constant for $p\rightarrow 0$.
However, data are fit very well by the relation 
\begin{equation}
\beta _{\text{a priori}}(p)=\frac{1}{1.25-0.82ln(p)}
\end{equation}
which would tend to zero as $p\rightarrow 0$ in conflict with our scaling
arguments. To resolve this dilemma we need to probe very small $p$. 


\section{The Limiting Case $p\rightarrow 0$}

We are interested in finding whether $\beta _{\text{a priori}}(p)$ tends
towards a constant as $p\rightarrow 0$. To do this using the above approach
is difficult, since we need a large number of cities to produce reliable
data for this regime. Extraction of this behaviour can however be achieved
by comparing simulations for different values of $n$, but fixed $np$. We
accomplish this by insisting that each instance has 4 cities on the pruned
tour. 4 city tours are chosen since they are the smallest for which it
matters in which order the cities are visited. This can be viewed as an
efficient way to simulate (approximately) the PTSP strategy with $p=\frac{4}{%
n}$.

\begin{figure}[tbp]
\resizebox{8.6cm}{5.5cm}{\includegraphics{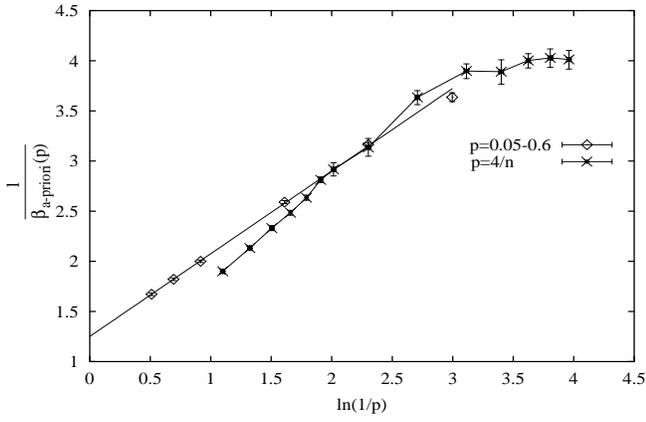}}
\caption{Reciprocal shift factors for a-priori tours (diamonds) compared to
estimates from 4 city tours (crosses),
$\frac{1}{\beta_{a-priori}(p)} \simeq \frac{n}{2 E\left( L_{\text{a priori}}^{\text{4 city}}\right)}$.
The 4 city tour data is optimized when each of the instances have 4 cities on
the pruned tour. The direct measurements do not appear to saturate
within the accessible range of $p$. The crosses show matching behaviour, with
saturation at larger $n$ corresponding to inaccessible $p$, suggesting that
$E(L_{\mbox{a priori}})= \beta_{0} \sqrt{\frac{n}{p}}$ for small $%
p$.}
\label{zerograph}
\end{figure}

Since we are considering the PTSP at fixed $np$, if $\beta _{\text{a priori}%
}(p)$ tends towards a (non-zero) constant as $p\rightarrow 0$ then we expect 
$E\left( L_{\text{a priori}}^{\text{4 city}}\right) /n$ to tend towards a
constant as $p\rightarrow 0$. Simulations in this regime were performed for $%
N=12-210$, \ with $100$ different random city configurations used for $N<30$%
, $20$ configurations for $N\leq 90$ and $10$ configurations for $N\geq 120$%
. Figure \ref{zerograph} shows a linear-log plot of $\frac{n}{2E\left( L_{%
\text{a priori}}^{\text{4 city}}\right) }$ against $\ln (n/4)\cong \ln (1/p)$%
. For small $n$ these results reasonably match the direct measurements of $%
\beta _{\text{a priori}}(p)$, shown for comparison. However, \ for
surprisingly large $n\sim 100$ which is beyond the range of our $\beta _{%
\text{a priori}}(p)$ data, our earlier proposal of scaling behaviour is
vindicated by $E\left( L_{\text{a priori}}^{\text{4 city}}\right) /n$
approaching a constant value. 
In summary we have 
\begin{equation}
E(L_{\text{a priori}})=\sqrt{\frac{n}{p}}\beta _{\text{a priori}}(p)
\label{a priori}
\end{equation}
where 
\begin{equation}
\beta _{\text{a priori}}(p)\left\{ 
\begin{array}{ll}
=\frac{1}{1.25-0.82\,ln(p)} & \;\;\;\;\;p>0.03 \\ 
=\beta _{0} & \;\;\;\;\;p<0.03.
\end{array}
\right.  \label{ba priori}
\end{equation}

\section{Notes on Algorithm Implementation}

\begin{figure}[tbp]
\resizebox{8.6cm}{4.3cm}{\includegraphics{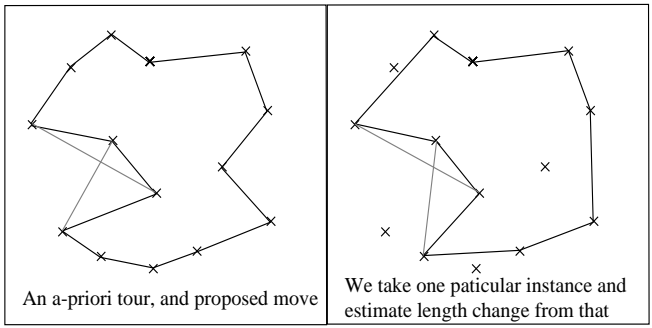}}
\caption{When estimating the expected length change due to a move, we
randomly generate instances. Only the cities that are nearest to the move
are needed to calculate the change in the pruned tour length.}
\label{diagram}
\end{figure}

We applied stochastic annealing to the PTSP using a combination of the 2-opt
and 1-shift move-sets\cite{lin1} established for the TSP. Both move-sets
work similarly to that which would be expected for the deterministic case.
The expected pruned tour length change for the move was estimated by
averaging the change in the tour length for a number of instances. For a
given instance it is not necessary to decide whether every city is present,
but only the set of cities closest to the move which determine the change in
the pruned tour length (see figure \ref{diagram}). For the PTSP, the
location of the nearest cities on the pruned tour to the move is determined
from a simple Poisson distribution.

\begin{figure}[tbp]
\resizebox{8.6cm}{6.5cm}{\includegraphics{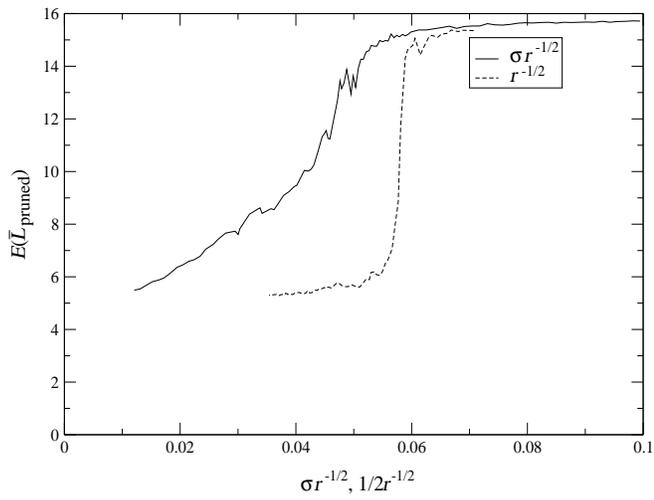}}
\caption{The expected pruned tour length for annealings when $r$
and $1/T=\frac{\protect\sqrt{r}}{\protect\sigma }$ are increased
monotonically. The sharp drop in the pruned tour length is seen when only $r$
is controlled, demonstrating that this ``freezes in'' imperfections in the
tour. \ The system was annealed at each value of the temperature and value
of $r$ for 50,000 Monte Carlo steps with $n=300$ and $p=0.1$.}
\label{glassy}
\end{figure}

When using stochastic optimization, the only variable over which we have
control is the sample size (the number of instances) $r$, whereas the effective
temperature $\frac{\sigma }{\sqrt{r}}$ also entails the standard deviation $%
\sigma $ of the pruned length change over instances. As shown in figure \ref
{glassy}, annealing by controlling $r$ alone exhibits a relatively sharp
transition in the expected pruned tour length. The rapid transition appears
to `freeze in' limitations in the tours found (analogous to defects in a
physical low temperature phase). By comparison we obtain a much smoother
change when $\frac{\sigma }{\sqrt{r}}$ is controlled.

The sharpness of the transition under control by $r$ is caused by the fact
that $\sigma $ may vary from move to move, and is on average lower when the
expected pruned tour length is less. The jump in the pruned tour length is
accompanied by a jump in $\sigma $ and hence the temperature. We suggest
that quite generally controlling $\frac{\sigma }{\sqrt{r}}$ gives a better
cooling schedule than focussing on $r$ alone.

\section{Conclusion}

We have shown that earlier incompatible ideas about the form of PTSP\ tours
especially at small $p$\cite{bertsimas1,bertsimas4,bartholdi} are resolved
by a new crossover scaling interpretation. \ The crossover scale corresponds
to a group of cities such that of order one will typically have to be
visited; \ below this scale the (optimsed) a priori PTSP tours resemble a
local sort whereas they are TSP-like on scales larger than the crossover.
Our computational results for the pruned tour length are summarised by eq. 
\ref{prunedlaw} and clearly support the crossover scaling.

Computationally the a priori tour length is more subtle than the pruned tour
length, \ although it does ultimately conform to expectations from crossover
scaling. We introduced 4-city tours to probe the behaviour of a priori tour
length down to very small $p$. As summarised by eq. \ref{a priori}, \ we
find a wide pre-asymptotic regime until \ recovering the expected crossover
scaling only for $p<0.03$. \ Understanding these anomalies in the a priori
tour length, \ and confirming them analytically, is left as a future
challenge.

We have shown stochastic annealing to be a robust and effective stochastic
optimization technique, \ taking the PTSP as a representative difficult
stochastic optimization problem. \ In this case it enabled us to obtain
representative results out to unprecedented problem sizes, \ which in turn
supported a whole new view of how the tours behave. \ Of relevance to wider
applications of stochastic optimmisation, \ we have seen that smoother
annealing can be obtained by directly controlling the effective temperature$%
\frac{\sigma }{\sqrt{r}}$\cite{fink1} rather than simply the bare depth of
sampling $r$ alone.

\hrulefill\ 

NEB would like to thank BP Amoco \& EPSRC for the support of a CASE award
during this research.

\appendix

\section*{The Length of a TSP Tour for small numbers of cities}

Numerical estimates of the length of a TSP tour for $n\le 10$ are given below

\bibliographystyle{prsty}
\bibliography{References}

\begin{thebibliography}{10}

\bibitem{fink1}
R.~C. Ball, T.~M. Fink, and N.~E. Bowler, submitted to Physical Review Letters,
  available at http://arXiv.org/abs/cond-mat/0301179 (unpublished).

\bibitem{jonsbraten}
T.~W. Jonsbraten, Journal of the operational research society {\bf 49},  811
  (1998).

\bibitem{robbins}
H. Robbins and S. Munro, The annals of mathematical statistics {\bf 22},  400
  (1951).

\bibitem{benveniste}
A. Benveniste, M. M\'etivier, and P. Priouret, {\em Adaptive algorithms and
  stochastic approximation} (Springer-Verlag, New York, 1990).

\bibitem{kushner1}
H.~J. Kushner and F.~J. V\'asquez, SIAM Journal on Control and Optimization
  {\bf 34},  712  (1996).

\bibitem{l'ecuyer}
P. L'Ecuyer and G. Yin, SIAM Journal on Optimization {\bf 8 No. 1},  217
  (1998).

\bibitem{kushner2}
H.~J. Kushner, SIAM Journal on Applied Mathematics {\bf 47},  169  (1987).

\bibitem{gong}
W.~B. Gong, Y.~C. Ho, and W. Zhai,  in {\em Proceedings of the 31st IEEE
  conference on decision and control} (IEEE, PO Box 1331, Piscataway, NJ,
  1992), pp.\ 795--802.

\bibitem{yan}
D. Yan and H. Mukai, SIAM journal on control and optimization {\bf 30 No. 3},
  594  (1992).

\bibitem{devroye}
L.~P. Devroye, IEEE Transactions on Information Theory {\bf 24},  142  (1978).

\bibitem{yakowitz}
S. Yakowitz and E. Lugosi, SIAM Journal on Scientific and Statistical Computing
  {\bf 11},  702  (1990).

\bibitem{andradottir}
S. Andrad\'ottir, SIAM Journal on Optimization {\bf 6 No. 2},  513  (1996).

\bibitem{haddock}
J. Haddock and J. Mittenthal, Computers and Industrial Engineering {\bf 22 No.
  4},  387  (1992).

\bibitem{bulgak}
A.~A. Bulgak and J.~L. Sanders,  in {\em Proceedings of the 1988 Winter
  Simulation Conference} (IEEE, PO Box 1331, Piscataway, NJ, 1988), pp.\
  684--690.

\bibitem{alrefaei}
M.~H. Alrefaei and S. Andrad\'ottir, Management Science {\bf 45 No. 5},  748
  (1999).

\bibitem{alkhamis}
T.~M.~A. Khamis, M.~A. Ahmed, and V.~K. Tuan, European Journal of Operational
  Research {\bf 116 No. 3},  530  (1999).

\bibitem{gutjahr}
W.~J. Gutjahr and G.~C. Pflug, Journal of global optimization {\bf 8},  1
  (1996).

\bibitem{gelfand}
S.~B. Gelfand and S.~K. Mitter, J. Optimization Theory and Applications {\bf
  62},  49  (1989).

\bibitem{rees}
S. Rees and R.~C. Ball, J. Phys. A {\bf 20},  1239  (1987).

\bibitem{nulton}
J.~D. Nulton and P. Salamon, Phys. Rev. A {\bf 37 No. 4},  1351  (1988).

\bibitem{salamon1}
P. Salamon, J.~D. Nulton, J.~R. Harland, J. Pedersen, G. Ruppiener, and L.
  Liao, Computer Physics Communications {\bf 49},  423  (1988).

\bibitem{bertsimas4}
D.~J. Bertsimas, P. Jaillet, and A.~R. Odoni, Operations Research {\bf 38 No.
  6},  1019  (1990).

\bibitem{beardwood}
J. Beardwood, J.~H. Halton, and J.~M. Hammersley, Proceedings of the Cambridge
  Philosophical Society {\bf 55},  299  (1959).

\bibitem{steele1}
J.~M. Steele, Annals of Probability {\bf 9},  365  (1981).

\bibitem{lee}
J. Lee and M.~Y. Choi, Phys. Rev. E {\bf 50},  R651  (1994).

\bibitem{jaillet1}
P. Jaillet, Ph.D. thesis, M.I.T., 1985.

\bibitem{jaillet2}
P. Jaillet, Operations research {\bf 36},  929  (1988).

\bibitem{bertsimas2}
D.~J. Bertsimas and L.~H. Howell, Eur. J. of Operational Research {\bf 65},  68
   (1993).

\bibitem{laporte}
G. Laporte, F.~V. Louveaux, and H. Mercure, Operations research {\bf 42 No. 3},
   543  (1994).

\bibitem{bertsimas1}
D.~J. Bertsimas, Ph.D. thesis, M.I.T., 1988.

\bibitem{rossi}
F.~A. Rossi and I. Gavioli,  in {\em Advanced school on stochastics in
  combinatorial optimization}, edited by G. Andreatta, F. Mason, and P.
  Serafini (World Scientific, Singapore, 1987), pp.\ 214--227.

\bibitem{bertsimas5}
D.~J. Bertsimas, P. Chervi, and M. Peterson, Transportation science {\bf 29 No.
  4},  342  (1995).

\bibitem{bartholdi}
J.~J. Bartholdi and L.~K. Blatzman, Operations Research Lett. {\bf 1},  121
  (1982).

\bibitem{lin1}
S. Lin, Bell Systems Technological Journal {\bf 44},  2245  (1965).

\end{thebibliography}

\begin{figure}[tbp]
\resizebox{8.6cm}{5.5cm}{\includegraphics{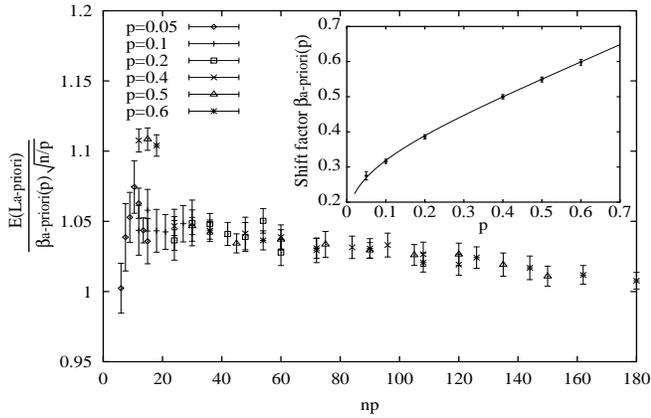}}
\caption{The master curve for the a priori tour length divided by $\protect%
\sqrt{\frac{n}{p}}\protect\beta _{\mbox{a priori}}(p)$. The shift factors,
inset, are expected to tend towards a constant for $p\rightarrow 0$. The
slight, but significant, deviation from linear suggests that this might not
be the case.}
\label{a priori graph}
\end{figure}

\begin{table*}[tbp]
{\centering 
\begin{tabular}{|c|c|c|c|}
\hline
Number of cities $n$ & Number of instances $I$ & Average tour length & $%
\sigma /\sqrt{I-1}$ \\ \hline\hline
2 & 100000 & 1.043 & 0.002 \\ \hline
3 & 100000 & 1.564 & 0.002 \\ \hline
4 & 5000 & 1.889 & 0.006 \\ \hline
5 & 5000 & 2.123 & 0.006 \\ \hline
6 & 5000 & 2.311 & 0.005 \\ \hline
7 & 5000 & 2.472 & 0.005 \\ \hline
8 & 5000 & 2.616 & 0.005 \\ \hline
9 & 5000 & 2.740 & 0.005 \\ \hline
10 & 5000 & 2.862 & 0.005 \\ \hline
\end{tabular}
}
\caption{The length of the optimal TSP tour for $n$ cities.}
\end{table*}

\end{document}